\documentclass[aps,showpacs,preprint,footinbib,preprintnumbers]{revtex4}

\def\ltap{\ \raise.3ex\hbox{$<$\kern-.75em\lower1ex\hbox{$\sim$}}\ }
\def\gtap{\ \raise.3ex\hbox{$>$\kern-.75em\lower1ex\hbox{$\sim$}}\ }
\def\gl{\ \raise.4ex\hbox{$>$\kern-.75em\lower1ex\hbox{$<$}}\ }

\usepackage[normalem]{ulem}  
\usepackage[dvips]{color} 
\usepackage{amsmath,amssymb}
\usepackage{booktabs}
\usepackage{mathrsfs}
\usepackage{graphicx}
\usepackage{multirow}

\renewcommand\sout{\bgroup \color{red} \ULdepth=-.5ex \ULset}

\newcommand{\Ex}[2]{\ifmmode{#1\times10^{#2}}\else{$#1\times10^{#2}$}\fi}

\begin{document}

\title{$\bar{\mathrm{D}}$ and $\mathrm{B}$ mesons in nuclear medium}
\author{S.~Yasui$^1$ and K.~Sudoh$^2$}
\affiliation{$^1$KEK Theory Center, Institute of Particle and Nuclear
Studies, High Energy Accelerator Research Organization, 1-1, Oho,
Ibaraki, 305-0801, Japan}
\affiliation{$^2$Nishogakusha University, 6-16, Sanbancho, Chiyoda,
Tokyo, 102-8336, Japan}
\date{\today}

\begin{abstract}
We discuss the mass modifications of $\bar{\mathrm{D}}$ and $\bar{\mathrm{D}}^{\ast}$ ($\mathrm{B}$ and $\mathrm{B}^{\ast}$) mesons in nuclear medium.
The heavy quark symmetry for $\bar{\mathrm{D}}$ and $\bar{\mathrm{D}}^{\ast}$ ($\mathrm{B}$ and $\mathrm{B}^{\ast}$) mesons is adopted, and the interaction between a $\bar{\mathrm{D}}$ or $\bar{\mathrm{D}}^{\ast}$  ($\mathrm{B}$ or $\mathrm{B}^{\ast}$) meson and a nucleon is supplied from the pion exchange.
We find the negative mass shifts 
for $\bar{\mathrm{D}}$ meson and 
 $\mathrm{B}$ meson, and hence that the $\bar{\mathrm{D}}$ and $\mathrm{B}$ mesons are bound in the nuclear medium.
As applications, we consider the atomic nuclei with a $\bar{\mathrm{D}}$ meson, $^{40}_{\bar{\mathrm{D}}}\mathrm{Ca}$ and $^{208}_{\bar{\mathrm{D}}}\mathrm{Pb}$, and investigate the energy levels of the $\bar{\mathrm{D}}$ meson in each nucleus.
We also discuss the mass shifts in the isospin asymmetric nuclear medium, and present the possible phenomenon about distribution of isospin density around a $\bar{\mathrm{D}}$ or $\mathrm{B}$ meson in nuclear medium.
We find that the mass shifts of $\bar{\mathrm{D}}^{\ast}$ and $\mathrm{B}^{\ast}$ mesons have large imaginary parts, which would prevent precise study of the energy levels of $\bar{\mathrm{D}}^{\ast}$ and $\mathrm{B}^{\ast}$ mesons in nuclei.
\end{abstract}

\maketitle

\section{Introduction}

Research of hadrons with charm and bottom flavors is one of the most interesting subjects in the present hadron and nuclear physics.
As motivated by  discoveries of new exotic hadrons with charm and bottom (X, Y, Z etc.) in experiments,
new types of hadrons, such as multi-quark or hadronic molecules, are investigated by many researchers \cite{Brambilla:2004wf,Swanson:2006st,Voloshin:2007dx,Nielsen:2009uh,Brambilla:2010cs}.
Among them, hadrons including $\bar{\mathrm{D}}$ or $\mathrm{B}$ meson are interesting objects not only in vacuum but also in nuclear medium.
Because a $\bar{\mathrm{D}}$ ($\mathrm{B}$) meson is composed with anti-charm quark $\bar{\mathrm{c}}$ (anti-bottom quark $\bar{\mathrm{b}}$) and a light quark $\mathrm{q}=\mathrm{u}, \mathrm{d}$,
there is no annihilation process of quark and antiquark pair 
  in nuclear medium.
The decay modes are only electromagnetic or weak processes.
Therefore, we expect to obtain precise information about the dynamics of $\bar{\mathrm{D}}$ ($\mathrm{B}$) meson in nuclear matter, such as their energy levels in atomic nuclei,
without being disturbed by the strong decay processes and absorption processes \cite{Tsushima:1998ru,Sibirtsev:1999js,Tsushima:2002cc}.
This nice property of $\bar{\mathrm{D}}$ and $\mathrm{B}$ mesons may be in contrast with their antiparticle states, $\mathrm{D}$ and $\bar{\mathrm{B}}$ mesons.
Because $\mathrm{D}$ and $\bar{\mathrm{B}}$ mesons in nuclear medium include light quark annihilation processes, 
the nonnegligible widths by the decays and absorptions make it difficult to study the dynamics of $\mathrm{D}$ and $\bar{\mathrm{B}}$ mesons in nuclear medium.

Study of $\bar{\mathrm{D}}$ and $\mathrm{B}$ mesons in nuclear matter will provide us with important information about (i) hadron-nucleon interaction, (ii) modification of hadrons in nuclear matter, and (iii) changes of nuclear medium caused by hadron as impurity.
As known in the light flavor mesons in nuclear medium, the mass shifts of vector mesons $\omega$, $\rho$, and $\phi$ give information of partial restoration of chiral symmetry breaking in nuclear matter \cite{Hayano:2008vn}.
It is also the case for $\bar{\mathrm{D}}$ and $\mathrm{B}$ mesons,
 because each of them contains a single light quark.
The properties of $\bar{\mathrm{D}}$ and $\mathrm{B}$ mesons are also concerned with the modifications of quarkonia in nuclear medium \cite{Klingl:1998sr,Song:2008bd,Morita:2010pd,Hayashigaki:2000es,Friman:2002fs}.
In early works \cite{Tsushima:1998ru,Sibirtsev:1999js,Tsushima:2002cc}, it was discussed that $\bar{\mathrm{D}}$ and $\mathrm{B}$ mesons are bound in atomic nuclei, such as $^{208}\mathrm{Pb}$.
The properties of $\bar{\mathrm{D}}$ mesons have been studied by several researchers with various theoretical approaches, such as QCD sum rule \cite{Klingl:1998sr,Song:2008bd,Morita:2010pd,Hayashigaki:2000es,Friman:2002fs,Hilger:2008jg,Hilger:2010zb,Wang:2011mj}, hadronic dynamics \cite{Mishra:2003se,Lutz:2005vx,Tolos:2007vh,Mishra:2008cd,Kumar:2010gb,JimenezTejero:2011fc,Kumar:2011ff,GarciaRecio:2011xt} and so on.
In the literature, the binding energy of a $\bar{\mathrm{D}}$ meson in normal nuclear matter is estimated to be around a few ten MeV.
In the present paper, we 
discuss the hadron dynamics by focusing on two 
important symmetries for heavy-light mesons; the heavy quark symmetry for a heavy quark and chiral symmetry for a light quark.

For the heavy quark symmetry, as a general property of QCD, the magnetic gluon couples to the heavy quark with a suppression factor $1/m_{\mathrm{Q}}$ with the heavy quark mass $m_{\mathrm{Q}}$.
It means that the spin of the heavy quark changes with a suppression by order of $1/m_{\mathrm{Q}}$, and hence it leads to the approximate mass degeneracy of pseudoscalar meson and vector meson in hadron mass spectrum \cite{Burdman:1992gh,Wise:1992hn,Yan:1992gz,Casalbuoni:1996pg,Manohar:2000dt}.
The mass splitting between  $\bar{\mathrm{D}}$ and $\bar{\mathrm{D}}^{\ast}$ is around 140 MeV.
The mass splitting between $\mathrm{B}$ and $\bar{\mathrm{B}}$ meson is around 45 MeV, hence the heavy quark symmetry becomes better for bottom flavor.
Indeed, these mass splittings are smaller than those in light flavor mesons; about 400 MeV between $\mathrm{K}$ and $\mathrm{K}^{\ast}$ mesons, and 630 MeV between $\pi$ and $\rho$ mesons.
Therefore, 
 due to the mass degeneracy in heavy-light mesons, we need to take int account both pseudoscalar mesons and vector mesons simultaneously as effective degrees of freedom.
The importance of the approximate degeneracy of $\bar{\mathrm{D}}$ and ${\mathrm{D}}^{\ast}$ ($\mathrm{B}$ and $\mathrm{B}^{\ast}$) mesons was already emphasized in studies of several exotic charm and bottom hadrons, such as two-body $\bar{\mathrm{D}}\mathrm{N}$ ($\mathrm{B}\mathrm{N}$) systems \cite{Yasui:2009bz,Yamaguchi:2011xb,Yamaguchi:2011qw,Carames:2012bd}, $\mathrm{Z}_{\mathrm{b}}$ mesons \cite{Ohkoda:2011vj}, $\mathrm{T}_{\mathrm{cc}}$ states \cite{Ohkoda:2012hv} \footnote{It is also discussed that $\mathrm{T}_{\mathrm{cc}}$ can be compact multi-quark states rather than extended hadronic molecular states. See Refs.~\cite{Lee:2007tn,Lee:2009rt} and the references therein.} and so on.
It will be shown in the text that the approximate mass degeneracy is important also for the dynamics of $\bar{\mathrm{D}}$ and ${\mathrm{D}}^{\ast}$ ($\mathrm{B}$ and $\mathrm{B}^{\ast}$) mesons in nuclear medium.

Another important symmetry is chiral symmetry.
The pion dynamics induced from the spontaneous chiral symmetry breaking in vacuum provides us with a key to understand the low energy hadron and nuclear physics \cite{Nambu:1961tp}.
In recent studies, it is discussed that the nuclear systems are almost governed by the pion dynamics following to the strong tensor force.
For example, the binding energies of nuclei are given about 80 \% by the pion exchange as shown in the few-body calculations \cite{Pieper:2001mp}.
The nuclear matter with infinite volume can also be described almost by the pion exchange between nucleons \cite{Kaiser:1997mw,Kaiser:2001jx,Kaiser:2001ra,Fritsch:2004nx,Fiorilla:2011qr,Hu:2009zza}.
Not only the normal nuclei, but also neutron-rich nuclei \cite{Myo:2007vm} and hypernuclei with strangeness \cite{Kaiser:2004fe,Kaiser:2007au,Finelli:2009tu}, are important objects in which the pion dynamics plays a dominant role.
The importance of the pion exchange was also emphasized for exotic charm and bottom hadrons in Refs.~\cite{Yasui:2009bz,Yamaguchi:2011xb,Yamaguchi:2011qw,Carames:2012bd,Ohkoda:2011vj,Ohkoda:2012hv} as above mentioned.
Especially, the various bound or resonant states of the two-body $\bar{\mathrm{D}}\mathrm{N}$ ($\mathrm{B}\mathrm{N}$) systems can be formed by the strong tensor force induced from the pion exchange between $\bar{\mathrm{D}}$ ($\mathrm{B}$) and $\mathrm{N}$ \cite{Yasui:2009bz,Yamaguchi:2011xb,Yamaguchi:2011qw,Carames:2012bd}.
Therefore, it is quite essential to consider the pion dynamics in order to discuss the properties of $\bar{\mathrm{D}}$ and $\bar{\mathrm{D}}^{\ast}$ ($\mathrm{B}$ and $\mathrm{B}^{\ast}$) mesons in nuclear medium.

In the present work, with respecting the heavy quark symmetry and chiral symmetry, we focus on the pion dynamics for $\bar{\mathrm{D}}$ and $\bar{\mathrm{D}}^{\ast}$ ($\mathrm{B}$ and $\mathrm{B}^{\ast}$) mesons in nuclear medium.
It will provide us with information about  (i) $\bar{\mathrm{D}}$- and $\bar{\mathrm{D}}^{\ast}$-nucleon ($\mathrm{B}$- and $\mathrm{B}^{\ast}$-nucleon) interaction and (ii) modification of $\bar{\mathrm{D}}$ and $\bar{\mathrm{D}}^{\ast}$ ($\mathrm{B}$ and $\mathrm{B}^{\ast}$) mesons  in nuclear matter.
We discuss shortly also (iii) changes of nuclear medium caused by $\bar{\mathrm{D}}$ and $\bar{\mathrm{D}}^{\ast}$ ($\mathrm{B}$ and $\mathrm{B}^{\ast}$) mesons as impurity.

This paper is organized as follows.
In Sec.~2, by following the description of the pion dynamics in nuclear matter \cite{Kaiser:1997mw,Kaiser:2001jx,Kaiser:2001ra} and hypernuclear matter \cite{Kaiser:2004fe,Kaiser:2007au,Finelli:2009tu},
we apply their approach to $\bar{\mathrm{D}}$ and $\bar{\mathrm{D}}^{\ast}$ ($\mathrm{B}$ and $\mathrm{B}^{\ast}$) mesons in nuclear matter.
We adopt the heavy quark symmetry for $\bar{\mathrm{D}}\bar{\mathrm{D}}^{\ast}\pi$ and $\bar{\mathrm{D}}^{\ast}\bar{\mathrm{D}}^{\ast}\pi$ ($\mathrm{B}\mathrm{B}^{\ast}\pi$ and $\mathrm{B}^{\ast}\mathrm{B}^{\ast}\pi$) vertices,
and obtain the self-energies of those particles in isospin symmetric nuclear matter.
We find that $\bar{\mathrm{D}}$ and $\mathrm{B}$ mesons have negative mass shifts 
 at the normal nuclear matter density, and hence they can be bound in nuclear matter.
In Sec.~3, we apply the results in the uniform nuclear matter to the study of the energy levels of $\bar{\mathrm{D}}$  meson in atomic nuclei.
We consider $^{40}_{\bar{\mathrm{D}}}\mathrm{Ca}$ and $^{208}_{\bar{\mathrm{D}}}\mathrm{Pb}$ nuclei, in which a $\bar{\mathrm{D}}$ meson is embedded in $^{40}\mathrm{Ca}$ and $^{208}\mathrm{Pb}$ nuclei.
We also discuss isospin asymmetric nuclear matter and find the upper and lower components of the isospin doublet of $\bar{\mathrm{D}}$ ($\mathrm{B}$) meson behave differently.
We discuss the possible phenomena about the distribution of isospin density around $\bar{\mathrm{D}}$ ($\mathrm{B}$) meson in nuclear medium.
In Sec.~4, we summarize our discussion and give some perspectives.

\section{$\bar{{\mathrm D}}$ and $\bar{{\mathrm D}}^{\ast}$ mesons in nuclear medium}

$\bar{\mathrm{D}}$ and $\bar{\mathrm{D}}^{\ast}$ ($\mathrm{B}$ and $\mathrm{B}^{\ast}$) mesons containing a heavy quark with a light antiquark have two different important symmetries; the heavy quark symmetry for a heavy quark and chiral symmetry for a light quark.
Both symmetries are incorporated in the effective theory for heavy-light mesons \cite{Burdman:1992gh,Wise:1992hn,Yan:1992gz,Casalbuoni:1996pg,Manohar:2000dt}.
The interaction Lagrangian with pion
 is given as
\begin{eqnarray}
{\cal L}_{\mathrm{int}}^{\mathrm{H}\mathrm{H}\pi} = -i \, \mathrm{Tr} \bar{H}_{a} v_{\mu} i {\cal V}_{ba}^{\mu} H_{b} + g_{\pi} \mathrm{Tr} \bar{H}_{a} H_{b} \gamma_{\mu} \gamma_{5} {\cal A}_{ba}^{\mu},
\end{eqnarray}
where we define the heavy-light meson field $H_{a}$ via pseudoscalar field $P_{a}$ ($\bar{\mathrm{D}}$ or $\mathrm{B}$ meson) and vector field $P_{a}^{\ast \, \mu}$ ($\bar{\mathrm{D}}^{\ast}$ or $\mathrm{B}^{\ast}$ meson),
\begin{eqnarray}
H_{a} = \frac{1+v\hspace{-0.5em}/}{2} \left[ P_{a}^{\ast \, \mu} \gamma_{\mu} + i P_{a} \gamma_{5} \right],
\end{eqnarray}
with the four dimensional vector $v$ with constraint $v^2=1$, isospin index $a$, and the Dirac indices $\mu=0, \cdots, 3$.
The vector and axial vector currents are ${\cal V}^{\mu} = \frac{i}{2}(\xi^{\dag} \partial^{\mu} \xi + \xi \partial^{\mu} \xi^{\dag})$ and ${\cal A}^{\mu} = \frac{i}{2}(\xi^{\dag} \partial^{\mu} \xi - \xi \partial^{\mu} \xi^{\dag})$ with $\xi = e^{i {\cal M}/f}$ and
\begin{eqnarray}
{\cal M} =
\left(
\begin{array}{cc}
 \frac{\pi^{0}}{\sqrt{2}} & \pi^{+} \\
 \pi^{-} & -\frac{\pi^{0}}{\sqrt{2}}
\end{array}
\right) = \frac{1}{\sqrt{2}} \vec{\pi} \!\cdot\! \vec{\tau}.
\end{eqnarray}
The coupling constant $g_{\pi}=0.59$ is determined from the $\bar{\mathrm{D}}\bar{\mathrm{D}}^{\ast}\pi$ decay width.
$f=132$ MeV is the pion decay constant.
The above vertex form is rewritten as
\begin{eqnarray}
{\cal L}_{\mathrm{int}}^{\mathrm{H}\mathrm{H}\pi} &=&
\frac{i}{f^2} P \, v^{\nu} ( \vec{\pi} \!\cdot\! \partial_{\nu} \vec{\pi} + i(\vec{\pi} \times \partial_{\nu} \vec{\pi}) \!\cdot\! \vec{\tau} ) P^{\dag} 
- \frac{i}{f^2} P^{\ast\,\mu} v^{\nu} ( \vec{\pi} \!\cdot\! \partial_{\nu} \vec{\pi} + i(\vec{\pi} \times \partial_{\nu} \vec{\pi}) \!\cdot\! \vec{\tau} ) P^{\ast \dag}_{\mu} \\ \nonumber
&&+ \frac{2ig_{\pi}}{f}  P \partial^{\mu} {\cal M} P^{\ast\,\dag}_{\mu} + \mathrm{h.c.} - \frac{2ig_{\pi}}{f} P_{a}^{\ast\, \alpha\, \dag} P_{b}^{\ast \, \beta} \partial^{\nu} {\cal M}_{ba} \epsilon_{\alpha \lambda \beta \nu} v^{\lambda}.
\end{eqnarray}
For nucleon parts, we define the $\mathrm{N}\mathrm{N}\pi$ vertex as
\begin{eqnarray}
{\cal L}_{\mathrm{int}}^{\mathrm{N}\mathrm{N}\pi} = -\frac{1}{4f^2} \bar{N} \gamma^{\mu} (\vec{\pi} \times \partial_{\mu} \vec{\pi}) \!\cdot\! \vec{\tau} N + \frac{g_{\mathrm{A}}}{2f} \bar{N} \gamma_{\mu} \gamma_{5} \partial^{\mu} \vec{\pi} \!\cdot\! \vec{\tau} N,
\end{eqnarray}
with
the coupling constant $g_{\mathrm{A}}=1.3$.
Then, the interaction between $\bar{\mathrm{D}}$ or $\bar{\mathrm{D}}^{\ast}$ and nucleon is given by the pion exchange \footnote{Following the procedure in Refs.~\cite{Kaiser:1997mw,Kaiser:2001jx,Kaiser:2001ra,Kaiser:2004fe,Kaiser:2007au,Finelli:2009tu}, we neglect the sigma terms
 and the $\Delta$ excitations.}.
The present formalism will be applied to $\mathrm{B}$ and $\mathrm{B}^{\ast}$ with better accuracy thanks to their heavy masses.

\begin{figure}[htbp]
\includegraphics[width=7cm,angle=-90]{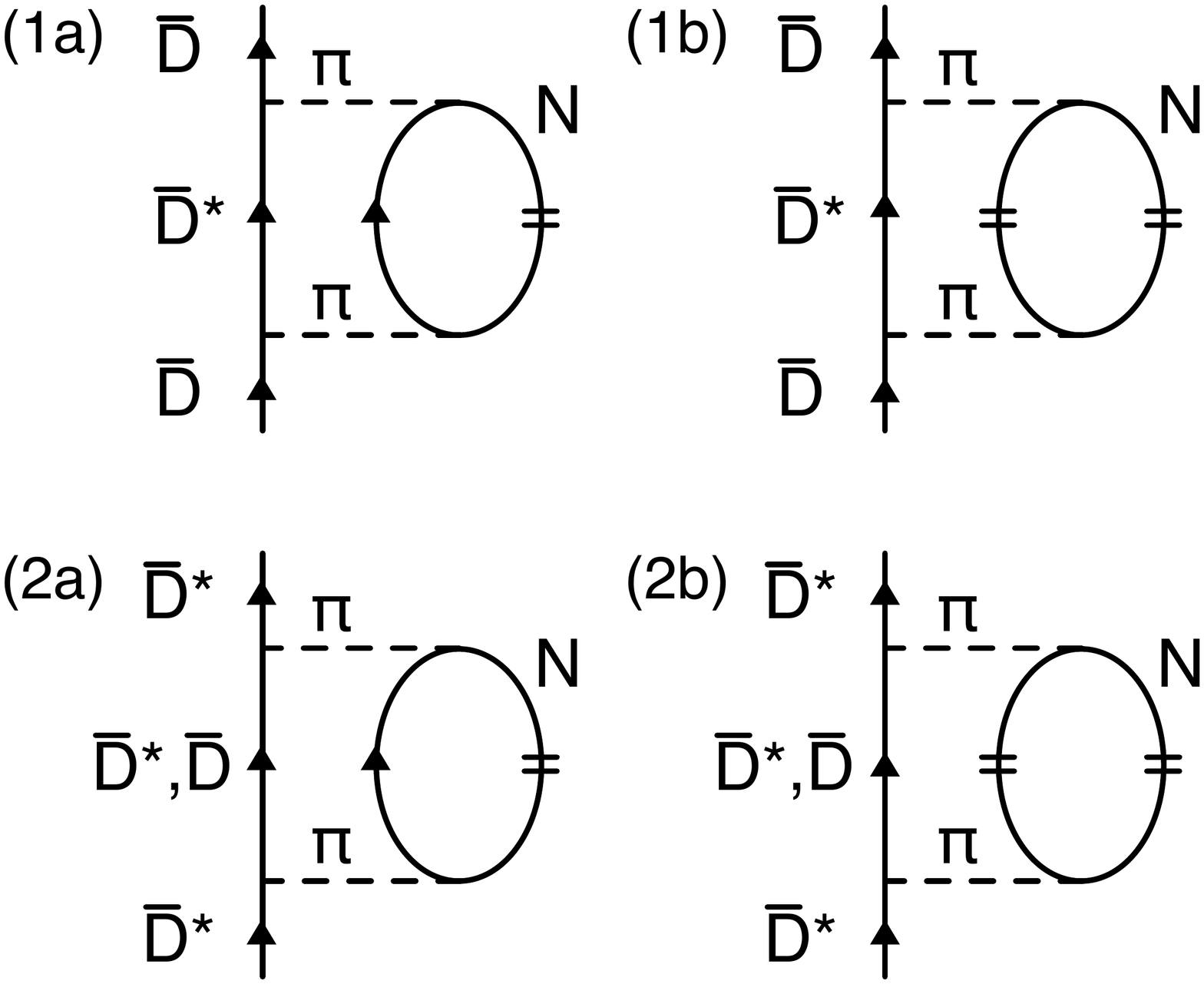}
\caption{The diagrams of the self-energies for $\bar{\mathrm{D}}$ (1a, b) and $\bar{\mathrm{D}}^{\ast}$ (2a, b) mesons. The horizontal double lines represent the second term in Eq.~(\ref{eq:propagator}). See the text for the details.}
\label{fig:Fig1}
\end{figure}

\begin{figure}[htbp]
\includegraphics[width=7cm,angle=-90]{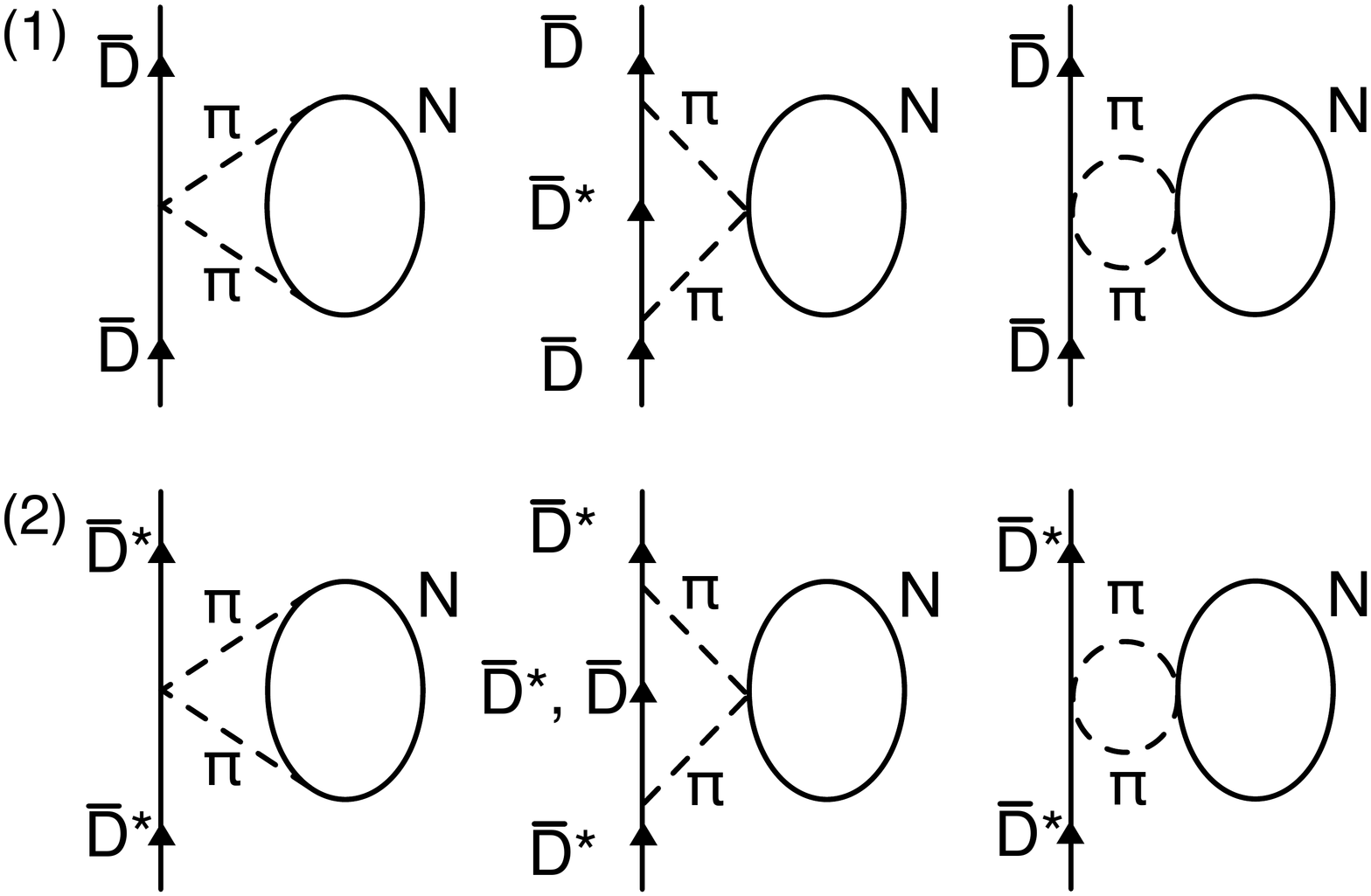}
\caption{Another diagrams of the self-energy for (1) $\bar{\mathrm{D}}$ meson and (2) $\bar{\mathrm{D}}^{\ast}$ meson.}
\label{fig:Fig2}
\end{figure}

Let us consider the self-energies of $\bar{\mathrm{D}}$ and $\bar{\mathrm{D}}^{\ast}$ mesons in uniform nuclear matter at zero temperature.
The leading contribution is given by the two pion exchange of which relevant diagrams are presented in Figs.~\ref{fig:Fig1} and \ref{fig:Fig2} \footnote{The diagrams in Fig.~\ref{fig:Fig2} do not appear in $\Lambda$ hyperon (isosinglet) in nuclear matter \cite{Kaiser:2004fe}.}.
However, the diagrams in Fig.~\ref{fig:Fig2} give zero contribution because of the cancellation of momenta at vertices or the antisymmetrization of isospin indices in isospin symmetric nuclear matter.
Therefore, we consider only the diagrams in Fig.~\ref{fig:Fig1}.
We may additionally consider the terms in which the two pion propagators are crossed.
However, such diagrams are suppressed by the inverse of the nucleon mass or heavy meson mass.
We neglect such higher contribution in the present discussion.
In the nucleon loop, we use the in-medium propagator for nucleon with mass $m_{\mathrm{N}}$ and momentum $p=(p_{0}, \vec{p}\,)$ in isospin symmetric nuclear matter with Fermi momentum $k_{\mathrm{F}}$,
\begin{eqnarray}
 (p\hspace{-0.4em}/+m_{\mathrm{N}}) \left\{ \frac{i}{p^{2}-m_{\mathrm{N}}^{2}+i\eta} - 2\pi \delta(p^{2}-m_{\mathrm{N}}^{2}) \theta(p_{0}) \theta(k_{\mathrm{F}}-|\vec{p}\,|) \right\} \mathbf{1}_{\mathrm{f}},
 \label{eq:propagator}
\end{eqnarray}
in which $\eta$ is an infinitely small positive number and $\mathbf{1}_{\mathrm{f}}$ is a $2\times2$ unit matrix in isospin space for proton and neutron \cite{Kaiser:1997mw,Kaiser:2001jx,Kaiser:2001ra,Kaiser:2004fe,Kaiser:2007au,Finelli:2009tu}.
The second term in Eq.~(\ref{eq:propagator}) is represented by the horizontal double line in Fig.~\ref{fig:Fig1}.
We assume, for simplicity, that the uniformity of nuclear matter does not change by the existence of the $\bar{\mathrm{D}}$ and $\bar{\mathrm{D}}^{\ast}$ mesons.
The discussion beyond this simplicity will be given later.
The nucleon mass in the propagator is set to be equal to the mass in vacuum.
The imaginary part is not adopted in the in-medium nucleon propagator 
 at zero temperature.
The Fermi momentum $k_{\mathrm{F}}$ is related to the nuclear matter density $n=2 k_{\mathrm{F}}^{3}/3\pi^{2}$ in isospin symmetric nuclear matter.

We give the expression of the self-energies for $\bar{\mathrm{D}}$ and $\bar{\mathrm{D}}^{\ast}$ mesons.
The same forms are applied to $\mathrm{B}$ and $\mathrm{B}^{\ast}$ mesons by replacing their masses \footnote{See also the discussions in Refs.~\cite{Yasui:2009bz,Yamaguchi:2011xb,Yamaguchi:2011qw}.}.
The self-energy of $\bar{\mathrm{D}}$ meson is
\begin{eqnarray}
\Sigma_{\bar{\mathrm{D}}}(k_{\mathrm{F}}) =
 \Sigma_{\bar{\mathrm{D}}}(k_{\mathrm{F}})^{(2\pi \bar{\mathrm{D}}^{\ast})} 
 + \Sigma_{\bar{\mathrm{D}}}(k_{\mathrm{F}})^{(2\pi \bar{\mathrm{D}}^{\ast})}_{\mathrm{Pauli}},
\end{eqnarray}
with
\begin{eqnarray}
\Sigma_{\bar{\mathrm{D}}}(k_{\mathrm{F}})^{(2\pi \bar{\mathrm{D}}^{\ast})} \!&=&\!
-\frac{6 g_{\pi}^{2} g_{A}^{2}}{f^{4}}
\int \frac{\mathrm{d}^{3} \vec{\ell}}{(2\pi)^{3}}
\int_{|\vec{p}_{1}| \le k_{\mathrm{F}}} \frac{\mathrm{d}^{3} \vec{p}_{1}}{(2\pi)^{3}}
\frac{m_{\mathrm{N}} |\vec{\ell}|^{4}}{\Delta^{2}+\frac{1}{2}(1+\frac{m_{\mathrm{N}}}{M}) |\vec{\ell}|^{\,2} + \vec{p}_{1} \!\cdot\! \vec{\ell} - i\eta} \hspace{0.2em} \frac{1}{(|\vec{\ell}|^{\,2} + m_{\pi}^{2})^{2}} \hspace{0.2em},  \nonumber \\
\end{eqnarray}
from 1a) in Fig.~\ref{fig:Fig1} and
\begin{eqnarray}
\Sigma_{\bar{\mathrm{D}}}(k_{\mathrm{F}})^{(2\pi \bar{\mathrm{D}}^{\ast})}_{\mathrm{Pauli}} \!&=&\!
\frac{6 g_{\pi}^{2} g_{A}^{2}}{f^{4}}
\int_{|\vec{p}_{2}| \le k_{\mathrm{F}}} \frac{\mathrm{d}^{3} \vec{p}_{2}}{(2\pi)^{3}}
\int_{|\vec{p}_{1}| \le k_{\mathrm{F}}} \frac{\mathrm{d}^{3} \vec{p}_{1}}{(2\pi)^{3}} \\ \nonumber
&& \frac{m_{\mathrm{N}} |\vec{p}_{1}-\vec{p}_{2}|^{4}}{\Delta^{2}+\frac{1}{2}(1+\frac{m_{\mathrm{N}}}{M}) |\vec{p}_{2}|^{\,2}+\frac{1}{2}(-1+\frac{m_{\mathrm{N}}}{M}) |\vec{p}_{1}|^{\,2} - \frac{m_{\mathrm{N}}}{M} \vec{p}_{1} \!\cdot\! \vec{p}_{2} - i\eta} \hspace{0.2em} \frac{1}{(|\vec{p}_{1}-\vec{p}_{2}|^{\,2} + m_{\pi}^{2})^{2}},
\end{eqnarray}
from 1b).
The self-energy of $\bar{\mathrm{D}}^{\ast}$ meson is
\begin{eqnarray}
\Sigma_{\bar{\mathrm{D}}^{\ast}}(k_{\mathrm{F}}) =
 \Sigma_{\bar{\mathrm{D}}^{\ast}}(k_{\mathrm{F}})^{(2\pi \bar{\mathrm{D}}^{\ast})}
 + \Sigma_{\bar{\mathrm{D}}^{\ast}}(k_{\mathrm{F}})^{(2\pi \bar{\mathrm{D}})} 
 + \Sigma_{\bar{\mathrm{D}}^{\ast}}(k_{\mathrm{F}})^{(2\pi \bar{\mathrm{D}}^{\ast})}_{\mathrm{Pauli}}
 + \Sigma_{\bar{\mathrm{D}}^{\ast}}(k_{\mathrm{F}})^{(2\pi \bar{\mathrm{D}})}_{\mathrm{Pauli}},
\end{eqnarray}
with
\begin{eqnarray}
\Sigma_{\bar{\mathrm{D}}^{\ast}}(k_{\mathrm{F}})^{(2\pi \bar{\mathrm{D}}^{\ast})} \!&=&\!
-\frac{4 g_{\pi}^{2} g_{A}^{2}}{f^{4}}
\int \frac{\mathrm{d}^{3} \vec{\ell}}{(2\pi)^{3}}
\int_{|\vec{p}_{1}| \le k_{\mathrm{F}}} \frac{\mathrm{d}^{3} \vec{p}_{1}}{(2\pi)^{3}}
\frac{m_{\mathrm{N}} |\vec{\ell}|^{4}}{\frac{1}{2}(1+\frac{m_{\mathrm{N}}}{M}) |\vec{\ell}|^{\,2} + \vec{p}_{1} \!\cdot\! \vec{\ell} - i\eta} \hspace{0.2em} \frac{1}{(|\vec{\ell}|^{\,2} + m_{\pi}^{2})^{2}} \hspace{0.2em}, \\
\Sigma_{\bar{\mathrm{D}}^{\ast}}(k_{\mathrm{F}})^{(2\pi \bar{\mathrm{D}})}  \!&=&\!
-\frac{2 g_{\pi}^{2} g_{A}^{2}}{f^{4}}
\int \frac{\mathrm{d}^{3} \vec{\ell}}{(2\pi)^{3}}
\int_{|\vec{p}_{1}| \le k_{\mathrm{F}}} \frac{\mathrm{d}^{3} \vec{p}_{1}}{(2\pi)^{3}}
\frac{m_{\mathrm{N}} |\vec{\ell}|^{4}}{-\Delta^{2}+\frac{1}{2}(1+\frac{m_{\mathrm{N}}}{M}) |\vec{\ell}|^{\,2} + \vec{p}_{1} \!\cdot\! \vec{\ell} - i\eta} \hspace{0.2em} \frac{1}{(|\vec{\ell}|^{\,2} + m_{\pi}^{2})^{2}} \hspace{0.2em}, \nonumber \\
\end{eqnarray}
from 2a) in the same figure, and
\begin{eqnarray}
\Sigma_{\bar{\mathrm{D}}^{\ast}}(k_{\mathrm{F}})^{(2\pi \bar{\mathrm{D}}^{\ast})}_{\mathrm{Pauli}} \!&=&\!
\frac{4 g_{\pi}^{2} g_{A}^{2}}{f^{4}}
\int_{|\vec{p}_{2}| \le k_{\mathrm{F}}} \frac{\mathrm{d}^{3} \vec{p}_{2}}{(2\pi)^{3}}
\int_{|\vec{p}_{1}| \le k_{\mathrm{F}}} \frac{\mathrm{d}^{3} \vec{p}_{1}}{(2\pi)^{3}} \\ \nonumber
&& \frac{m_{\mathrm{N}} |\vec{p}_{1}-\vec{p}_{2}|^{4}}{\frac{1}{2}(1+\frac{m_{\mathrm{N}}}{M}) |\vec{p}_{2}|^{\,2}+\frac{1}{2}(-1+\frac{m_{\mathrm{N}}}{M}) |\vec{p}_{1}|^{\,2} - \frac{m_{\mathrm{N}}}{M} \vec{p}_{1} \!\cdot\! \vec{p}_{2} - i\eta} \hspace{0.2em} \frac{1}{(|\vec{p}_{1}-\vec{p}_{2}|^{\,2} + m_{\pi}^{2})^{2}} \hspace{0.2em}, \\
\Sigma_{\bar{\mathrm{D}}^{\ast}}(k_{\mathrm{F}})^{(2\pi \bar{\mathrm{D}})}_{\mathrm{Pauli}} \!&=&\!
\frac{2 g_{\pi}^{2} g_{A}^{2}}{f^{4}}
\int_{|\vec{p}_{2}| \le k_{\mathrm{F}}} \frac{\mathrm{d}^{3} \vec{p}_{2}}{(2\pi)^{3}}
\int_{|\vec{p}_{1}| \le k_{\mathrm{F}}} \frac{\mathrm{d}^{3} \vec{p}_{1}}{(2\pi)^{3}} \\ \nonumber
&& \frac{m_{\mathrm{N}} |\vec{p}_{1}-\vec{p}_{2}|^{4}}{-\Delta^{2}+\frac{1}{2}(1+\frac{m_{\mathrm{N}}}{M}) |\vec{p}_{2}|^{\,2}+\frac{1}{2}(-1+\frac{m_{\mathrm{N}}}{M}) |\vec{p}_{1}|^{\,2} - \frac{m_{\mathrm{N}}}{M} \vec{p}_{1} \!\cdot\! \vec{p}_{2} - i\eta} \hspace{0.2em} \frac{1}{(|\vec{p}_{1}-\vec{p}_{2}|^{\,2} + m_{\pi}^{2})^{2}} \hspace{0.2em},
\end{eqnarray}
from 2b).
We note that there is no $\bar{\mathrm{D}}$ state in the intermediate state in the self-energy of $\bar{\mathrm{D}}$ meson, because the $\bar{\mathrm{D}}\bar{\mathrm{D}}\pi$ vertex is forbidden due to parity conservation.
We define $\Delta^{2}/m_{\mathrm{N}} = m_{\bar{\mathrm{D}}^{\ast}} - m_{\bar{\mathrm{D}}}$ as a mass difference between $\bar{\mathrm{D}}$ and $\bar{\mathrm{D}}^{\ast}$ mesons, and $M = (m_{\bar{\mathrm{D}}} + 3 m_{\bar{\mathrm{D}}^{\ast}})/4$ as an averaged meson mass.
In the integrals, $\vec{\ell}$, $\vec{p}_1$, and $\vec{p}_2$ are three-dimensional momenta for pion ($\vec{\ell}$) and nucleon ($\vec{p}_1$, $\vec{p}_2 = \vec{p}_1 + \vec{\ell}$) in the loops.
To obtain the above equations, we first accomplish the four-dimensional integral and then expand by the inverse of the nucleon mass and $\bar{\mathrm{D}}$ and $\bar{\mathrm{D}}^{\ast}$ meson masses.
We consider only the leading terms in the expansion.
It is known that the approximation works not only in normal nuclear matter but also in hypernuclear matter \cite{Kaiser:1997mw,Kaiser:2001jx,Kaiser:2001ra,Kaiser:2004fe,Kaiser:2007au,Finelli:2009tu}. 
It will work well also for charm and bottom systems because the masses are heavier than those of nucleons or hyperons.

Because the integrals by $\vec{\ell}$ diverges, the regularization should be performed.
According to the procedure in Refs.~\cite{Kaiser:2004fe,Kaiser:2007au}, we adopt the following regularization for the real parts;
\begin{eqnarray}
&& \mathrm{P} \int \frac{\mathrm{d}^{3} \vec{\ell}}{(2\pi)^{3}}
\frac{|\vec{\ell}|^{4}}{x+\frac{1}{2}(1+\frac{m_{\mathrm{N}}}{M}) |\vec{\ell}|^{\,2} + \vec{p}_{1} \!\cdot\! \vec{\ell} - i\eta} \hspace{0.2em} \frac{1}{(|\vec{\ell}|^{\,2} + m_{\pi}^{2})^{2}} \nonumber \\
&\rightarrow &
\mathrm{P} \int \frac{\mathrm{d}^{3} \vec{\ell}}{(2\pi)^{3}}
\frac{|\vec{\ell}|^{4}}{x+\frac{1}{2}(1+\frac{m_{\mathrm{N}}}{M}) |\vec{\ell}|^{\,2} + \vec{p}_{1} \!\cdot\! \vec{\ell} - i\eta} \hspace{0.2em} \frac{1}{(|\vec{\ell}|^{\,2} + m_{\pi}^{2})^{2}}
- \int \frac{\mathrm{d}^{3} \vec{\ell}}{(2\pi)^{3}}
\frac{1}{\frac{1}{2}(1+\frac{m_{\mathrm{N}}}{M}) |\vec{\ell}|^{\,2}} \nonumber \\
&+&
\int_{|\vec{\ell}| \le \Lambda} \frac{\mathrm{d}^{3} \vec{\ell}}{(2\pi)^{3}}
\frac{1}{\frac{1}{2}(1+\frac{m_{\mathrm{N}}}{M}) |\vec{\ell}|^{\,2}},
\end{eqnarray}
with $x=0$, $\pm \Delta^{2}$.
Here, $\mathrm{P}$ stands for the principal integration.
The sum of the first and second integrals becomes finite in the integration at $|\vec{\ell}| \rightarrow \infty$, while the last term becomes finite by introducing three-dimensional momentum cutoff parameter $\Lambda$.
Thus, we succeed to separate the cutoff-independent (first and second) term and -dependent (third) term as discussed in Refs.~\cite{Kaiser:1997mw,Kaiser:2001jx,Kaiser:2001ra,Kaiser:2004fe,Kaiser:2007au,Finelli:2009tu}.
The three-dimensional momentum cutoff parameters are set to be
$\Lambda_{\bar{\mathrm{D}}} = 1.27 \Lambda_{\Lambda}$ and
$\Lambda_{\mathrm{B}} = 1.22 \Lambda_{\Lambda}$
for $\bar{\mathrm{D}}$ and $\mathrm{B}$ mesons, respectively, with $\Lambda_{\Lambda}=700$ MeV for $\Lambda$ hyperon in nuclear matter \cite{Kaiser:2004fe}.
The ratios of $\Lambda_{\bar{\mathrm{D}}}/\Lambda_{\Lambda}$ or $\Lambda_{\mathrm{B}}/\Lambda_{\Lambda}$ are estimated by assuming that the cutoff momentum is proportional to the inverse of the size of hadron; $\Lambda_{\mathrm{P}}/\Lambda_{\Lambda}=r_{\Lambda}/r_{\mathrm{P}}$ with mean-matter-radii $r_{\mathrm{P}}$ ($r_{\Lambda}$) of $\mathrm{P}=\bar{\mathrm{D}}$, $\mathrm{B}$ meson ($\Lambda$ hyperon) which is determined from the quark model \cite{Yasui:2009bz}.
This relation is satisfied well
 for the ratio of $\Lambda$ and nucleon.
Indeed, we find that $\Lambda_{\mathrm{N}}/\Lambda_{\Lambda}=650\,\mathrm{MeV}/700\,\mathrm{MeV}=0.93$ 
to reproduce the properties of (hyper)nuclear matter \cite{Kaiser:1997mw,Kaiser:2001jx,Kaiser:2001ra,Kaiser:2004fe,Kaiser:2007au,Finelli:2009tu} is comparable with $r_{\Lambda}/r_{\mathrm{N}}=0.94$ from the mean-matter-radius estimated in the quark model.
In the numerical calculations, we use the following mass values; $m_{\mathrm{N}}=938.9$ MeV, $m_{\pi}=138.0$ MeV, $m_{\bar{\mathrm{D}}}=1867.2$ MeV, $m_{\bar{\mathrm{D}}^{\ast}}=2008.6$ MeV, $m_{\mathrm{B}}=5279.3$ MeV, and $m_{\mathrm{B}^{\ast}}=5325.1$ MeV.

Before discussing the numerical results, we remark that
the imaginary part of the $\bar{\mathrm{D}}$ ($\mathrm{B}$) meson self-energy is zero at low density.
Because there is a mass gap between $\bar{\mathrm{D}}$ and $\bar{\mathrm{D}}^{\ast}$ ($\mathrm{B}$ and $\mathrm{B}^{\ast}$) mesons in the intermediate states in 1a,b) in Fig.~\ref{fig:Fig1}.
The imaginary parts are absent for $n < n_{\mathrm{cr}} = 2(k_{\mathrm{F}}^{\mathrm{cr}})^{2}/3\pi^{2}$, where the critical Fermi momentum is given by
\begin{eqnarray}
k_{\mathrm{F}}^{\mathrm{cr}} = \sqrt{2\left( 1+\frac{m_{\mathrm{N}}}{M} \right)} \Delta.
\label{eq:critical_Fermi}
\end{eqnarray}
This can be checked, because the functions in the momentum integrals for the $\bar{\mathrm{D}}$ ($\mathrm{B}$) meson self-energy have no pole for $k_{\mathrm{F}} < k_{\mathrm{F}}^{\mathrm{cr}}$.
The imaginary part appears only at high density, $n > n_{\mathrm{cr}}$.
As numerical values, we obtain the critical densities $n_{\mathrm{cr}} = 2.2$ $\mathrm{fm}^{-3}$ for $\bar{\mathrm{D}}$ meson and $n_{\mathrm{cr}} = 0.28$ $\mathrm{fm}^{-3}$ for $\mathrm{B}$ meson.
In both cases, the critical densities are larger than the normal nuclear matter density $n_{0}=0.17$ $\mathrm{fm}^{-3}$.
Because the present formalism cannot be applied at high densities, we do not need to consider such high density state and ignore the imaginary part of the $\bar{\mathrm{D}}$ ($\mathrm{B}$) meson self-energy. 
On the other hand, 
 the imaginary parts in the self-energies of $\bar{\mathrm{D}}^{\ast}$ and $\mathrm{B}^{\ast}$ mesons exist even at low densities.
The value of the imaginary part is independent of the cutoff parameter.

\begin{figure}[htbp]
\includegraphics[width=11cm]{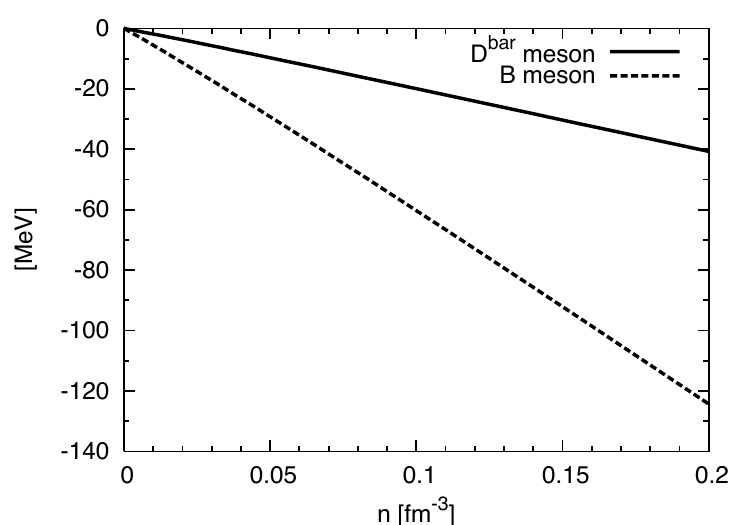}
\caption{The real part of the self-energies for $\bar{\mathrm{D}}$ and $\mathrm{B}$ mesons.}
\label{fig:D_density}
\end{figure}

\begin{figure}[htbp]
\includegraphics[width=11cm]{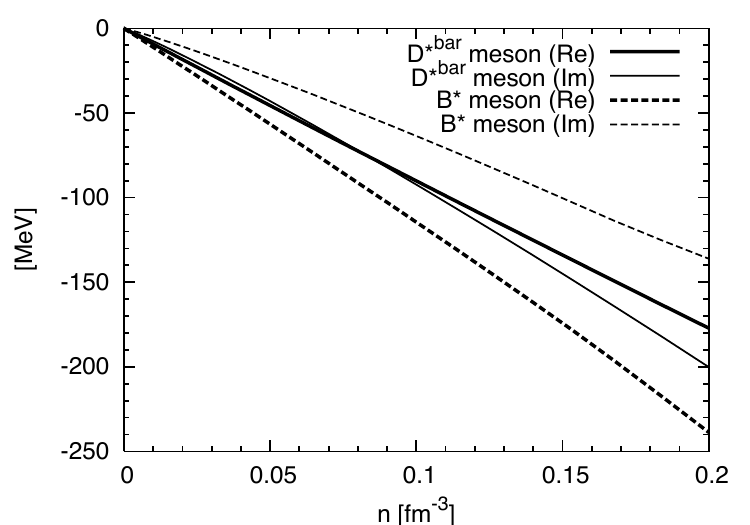}
\caption{The real and imaginary parts of the self-energies for $\bar{\mathrm{D}}^{\ast}$ and $\mathrm{B}^{\ast}$ mesons.}
\label{fig:Dstar_density}
\end{figure}

We present the numerical results for 
$\bar{\mathrm{D}}$ and $\bar{\mathrm{D}}^{\ast}$ ($\mathrm{B}$ and $\mathrm{B}^{\ast}$) mesons in the isospin symmetric nuclear matter. 
The self-energies of $\bar{\mathrm{D}}$ and $\bar{\mathrm{D}}^{\ast}$ ($\mathrm{B}$ and $\mathrm{B}^{\ast}$) mesons are plotted as functions of nuclear number density in Figs.~\ref{fig:D_density} and \ref{fig:Dstar_density}, respectively.
We obtain $-35.1$ MeV for $\bar{\mathrm{D}}$ meson and $-106.9$ MeV for $\mathrm{B}$ meson at the normal nuclear matter density $n_{0}$.
As mentioned before, there is no imaginary part.
On the other hand, we obtain $-153.9-i170.1$ MeV for $\bar{\mathrm{D}}^{\ast}$ meson and $-203.3-i117.5$ MeV for $\mathrm{B}^{\ast}$ meson at the normal nuclear matter density $n_{0}$.
Therefore, these states are unstable in nuclear medium due to the large decay widths, though they are long-lived in vacuum with small decay widths (less than a few MeV for $\bar{\mathrm{D}}^{\ast}$ meson and zero for $\mathrm{B}^{\ast}$ meson in strong interaction). 

The negative mass shifts of $\bar{\mathrm{D}}$ and $\bar{\mathrm{D}}^{\ast}$ ($\mathrm{B}$ and $\mathrm{B}^{\ast}$) mesons indicate that they can be bound in nuclear matter.
This result will be reasonable, because
the interaction between 
 a $\bar{\mathrm{D}}$ ($\mathrm{B}$) meson and a nucleon $\mathrm{N}$ 
can be attractive 
 by the pion exchange as discussed in Refs.~\cite{Yasui:2009bz,Yamaguchi:2011xb,Yamaguchi:2011qw,Carames:2012bd}.
There, not only the two-body $\bar{\mathrm{D}}\mathrm{N}$ ($\mathrm{B}\mathrm{N}$) systems, but also the possible three-body $\bar{\mathrm{D}}\mathrm{N}\mathrm{N}$ ($\mathrm{B}\mathrm{N}\mathrm{N}$) systems have been discussed \cite{Yasui:2009bz}.
Therefore, we expect that the $\bar{\mathrm{D}}$ ($\mathrm{B}$) meson and nucleon systems have a rich variety from small baryon numbers to infinitely large baryon numbers.

\section{Discussions}

\subsection{Energy levels of a $\bar{\mathrm{D}}$ meson in atomic nuclei}

\begin{figure}[htbp]
\includegraphics[width=9cm,angle=-90]{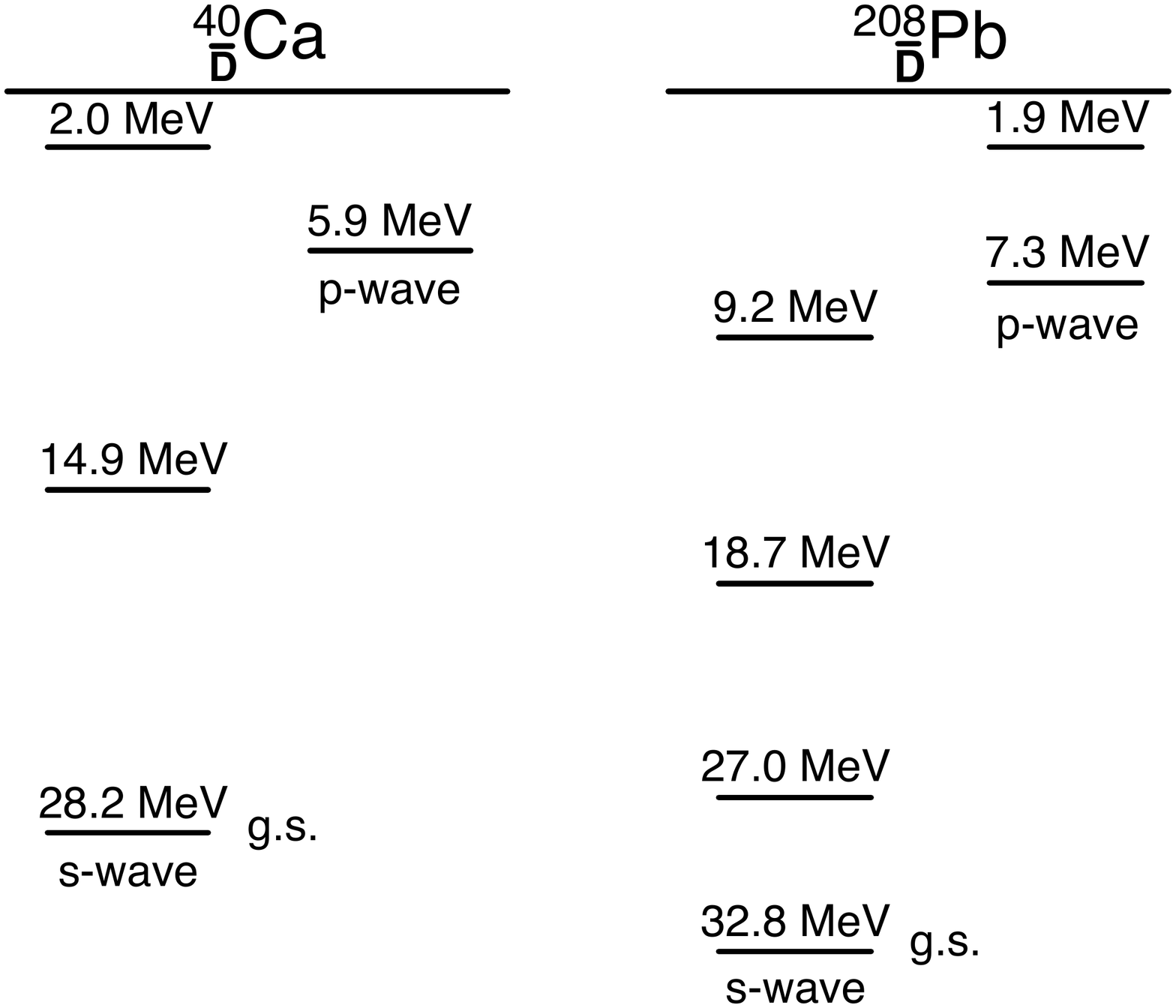}
\caption{The energy levels of $\bar{\mathrm{D}}$ meson in $^{40}_{\bar{\mathrm{D}}}\mathrm{Ca}$ and $^{208}_{\bar{\mathrm{D}}}\mathrm{Pb}$.}
\label{fig:Fig3}
\end{figure}

We consider the atomic nuclei containing a $\bar{\mathrm{D}}$ meson.
The potential of $\bar{\mathrm{D}}$ meson in the nucleus can be given as the real part of the self-energy \footnote{Because we give here qualitative discussions, we neglect the momentum-dependence in the self-energy.}.
We assume that the properties (density distribution, shape, and so on) of the core nucleus do not change by the existence of $\bar{\mathrm{D}}$ meson and adopt the Woods-Saxson-type potential for $\bar{\mathrm{D}}$ meson
\begin{eqnarray}
V(r;A) = \frac{V_{0}}{1+e^{(r-R_A)/a }},
\label{eq:distribution}
\end{eqnarray}
with nucleus radius $R_{A} = r_{0} A^{1/3}$ with baryon number $A$.
The parameters are $r_{0}=1.27$ fm, $a=0.67$ fm from the empirical data for stable nuclei (see for example Ref.~\cite{BohrMottelson}).
$V_{0}$ is the potential at the center of the nucleus.
From the results of the self-energies in the previous section, we use $V_{0}=-35.1$ MeV for $\bar{\mathrm{D}}$ meson.
We solve numerically the Klein-Gordon equation in the non-relativistic approximation and obtain the energy levels of $\bar{\mathrm{D}}$ meson in $^{40}_{\bar{\mathrm{D}}}\mathrm{Ca}$ and $^{208}_{\bar{\mathrm{D}}}\mathrm{Pb}$, in which $\bar{\mathrm{D}}$ meson is embedded in $^{40}\mathrm{Ca}$ and $^{208}\mathrm{Pb}$ nuclei, respectively \footnote{For simplicity it is assumed that the proton and neutron numbers in $^{208}\mathrm{Pb}$ is the same. For more realistic discussion, we need to consider the difference of the proton and neutron numbers.}.
As shown in Fig.~\ref{fig:Fig3}, several s- and p-wave orbitals of $\bar{\mathrm{D}}$ meson inside the nucleus exist.
The binding energies for the ground states (g.s.) of $\bar{\mathrm{D}}$ meson are 28.2 MeV for $^{40}_{\bar{\mathrm{D}}}\mathrm{Ca}$ and 32.8 MeV for $^{208}_{\bar{\mathrm{D}}}\mathrm{Pb}$.
The latter binding energy is close to the absolute value of the self-energy -35.1 MeV of $\bar{\mathrm{D}}$ meson in nuclear matter with infinite volume.

The energy levels for $\bar{\mathrm{D}}^{\ast}$  meson in atomic nuclei may be discussed similarly.
In this case, we need to introduce the complex potential with a finite imaginary part.
The complex potential can be regarded as the self-energy of $\bar{\mathrm{D}}^{\ast}$ meson in the previous section.
However, the imaginary part
 is so large ($\sim 100$ MeV) that the fine energy splittings ($\sim$ a few MeV) between the orbitals of $\bar{\mathrm{D}}^{\ast}$ meson in nuclei will not be seen clearly.
In comparison with $\bar{\mathrm{D}}$ mesons, therefore, $\bar{\mathrm{D}}^{\ast}$ mesons are not so useful objects for the study of energy levels in nuclei.

So far we have considered $\bar{\mathrm{D}}$ 
mesons.
Similar discussion may be applied to their antiparticle states, namely $\mathrm{D}$ 
 mesons.
In fact the only modification is just a change of the sign of the vertices of $\bar{\mathrm{D}} \bar{\mathrm{D}}^{\ast} \pi$ and $\bar{\mathrm{D}}^{\ast} \bar{\mathrm{D}}^{\ast} \pi$.
However, there appear additional strong decay processes ($\mathrm{D} \mathrm{N} \rightarrow \pi \Sigma_{\mathrm{c}}$, $\pi \Sigma_{\mathrm{c}}^{\ast} $, and so on) as well as two-body absorption processes ($\mathrm{D} \mathrm{N} \mathrm{N} \rightarrow \Lambda_{\mathrm{c}} \mathrm{N}$, $\Sigma_{\mathrm{c}} \mathrm{N}$, $\Sigma_{\mathrm{c}}^{\ast}  \mathrm{N}$, $\Lambda_{\mathrm{c}}^{\ast} \mathrm{N}$, and so on).
Those channels must make the analysis be complicated and give finite imaginary parts of the self-energy of $\mathrm{D}$
meson \cite{Bayar:2012dd}.
The similar situation will be also the case for $\bar{\mathrm{B}}$ meson.
Therefore, we claim again that $\bar{\mathrm{D}}$ and $\mathrm{B}$ mesons are unique hadrons to study the fine energy splittings of $\bar{\mathrm{D}}$ and $\mathrm{B}$ meson orbitals in nuclei.

\subsection{Isospin asymmetric nuclear matter}

In isospin asymmetric nuclear matter,
the Fermi surface is unbalanced due to the different number densities of protons and neutrons.
We introduce the Fermi momenta $k_{\mathrm{F}}^{\mathrm{p}}$ and $k_{\mathrm{F}}^{\mathrm{n}}$ for protons and neutrons, respectively.
Then, the step function in the in-medium nucleon propagator in Eq.~(\ref{eq:propagator}) is modified as
\begin{eqnarray}
\theta(k_{\mathrm{F}}-|\vec{p}\,|) \mathbf{1}_{\mathrm{f}} \rightarrow  \theta(k_{\mathrm{F}}^{\mathrm{p}}-|\vec{p}\,|) \frac{1+\tau_{3}}{2} + \theta(k_{\mathrm{F}}^{\mathrm{n}}-|\vec{p}\,|) \frac{1-\tau_{3}}{2},
\label{eq:propagator_asym}
\end{eqnarray}
where $\mathbf{1}_{\mathrm{f}}$ is a $2 \times 2$ unit matrix in the isospin space, and $(1 \pm \tau_{3})/2$ are projection operators for protons and neutrons, respectively \cite{Kaiser:2001jx}.
It is convenient to introduce the asymmetry parameter $\delta$ by giving $k_{\mathrm{F}}^{\mathrm{p}}$ and $k_{\mathrm{F}}^{\mathrm{n}}$ as
\begin{eqnarray}
k_{\mathrm{F}}^{\mathrm{p},\mathrm{n}} = k_{\mathrm{F}} (1 \mp \delta)^{1/3},
\end{eqnarray}
where $\delta$ is equal to $(n_{\mathrm{n}}-n_{\mathrm{p}})/(n_{\mathrm{n}}+n_{\mathrm{p}})$ with proton or neutron number density $n_{\mathrm{p},\mathrm{n}}$.
$\delta =0$ is the case of isospin symmetric nuclear matter, and $\delta = 1$ is the case of neutron matter.
The total nuclear matter density is $n = n_{\mathrm{n}}+n_{\mathrm{p}}$ with $n=2 k_{\mathrm{F}}^{3}/3\pi^{2}$ and $n_{\mathrm{p},\mathrm{n}} = (k_{\mathrm{F}}^{\mathrm{p},\mathrm{n}})^{3}/3\pi^{2}$.

\begin{figure}[htbp]
 \begin{minipage}{0.5\hsize}
  \begin{center}
   \includegraphics[width=11cm]{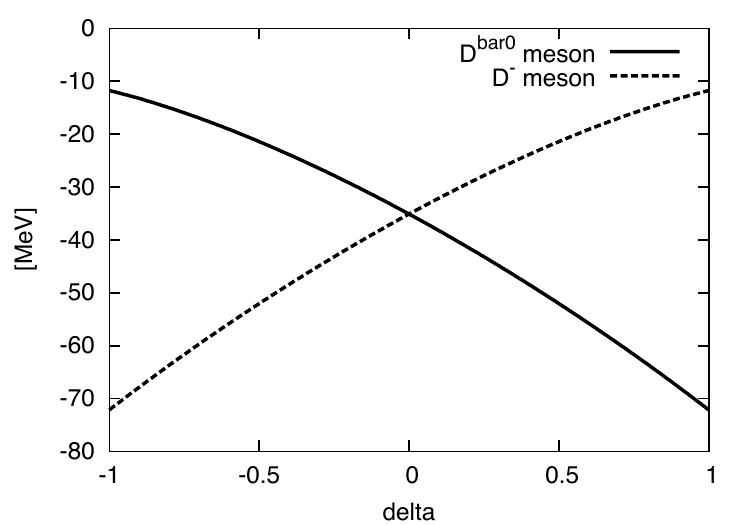}
  \end{center}
 \end{minipage}
 \begin{minipage}{0.5\hsize}
  \begin{center}
   \includegraphics[width=11cm]{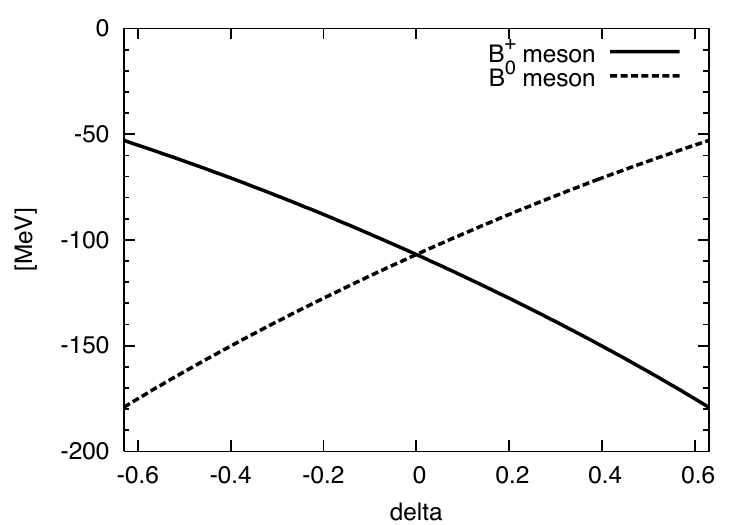}
  \end{center}
 \end{minipage}
\caption{The mass splittings of $\bar{\mathrm{D}} = (\bar{\mathrm{D}}^{0}, \mathrm{D}^{-})$ (top) and $\mathrm{B} = (\mathrm{B}^{+}, \mathrm{B}^{0})$ (bottom) at the normal nuclear matter density $n_{0} = 0.17$ $\mathrm{fm}^{-3}$ as functions of the asymmetry parameter $\delta$.}
  \label{fig:D_asymmetry}
\end{figure}

\begin{figure}[htbp]
\includegraphics[width=11cm]{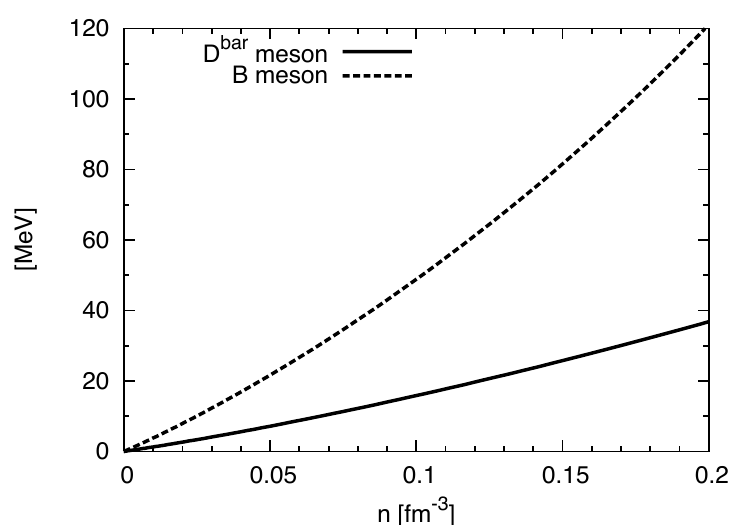}
\caption{$\sigma_{\bar{\mathrm{D}}}$ and $\sigma_{\mathrm{B}}$ as functions of nuclear matter density $n$.}
\label{fig:sigma_density_asy}
\end{figure}

When the nuclear matter is isospin asymmetric ($\delta \neq 0$), we observe mass splitting of up and down components in isospin doublet of $\bar{\mathrm{D}} = (\bar{\mathrm{D}}^{0}, \mathrm{D}^{-})$ or $\mathrm{B} = (\mathrm{B}^{+}, \mathrm{B}^{0})$.
We remark that the mass splitting of the isospin doublet is zero in the isospin symmetric nuclear matter. 
We obtain the self-energies of the upper and lower components in the isospin doublet $\bar{\mathrm{D}}$ or $\mathrm{B}$ by using the in-medium nucleon propagator in Eq.~(\ref{eq:propagator_asym}) \footnote{Following the discussions in the previous section, we use the self-energy of the diagrams of Fig.~\ref{fig:Fig1}.
According to Ref.~\cite{Dote:2002db}, one can expect that the current discussion will not change qualitatively, when 
additional diagrams are included.}.
In Fig.~\ref{fig:D_asymmetry}, we present the self-energies of $\bar{\mathrm{D}}^{0}$ and $\mathrm{D}^{-}$ mesons as functions of the asymmetry parameter $\delta$ at the normal nuclear matter density $n_{\mathrm{p}}+n_{\mathrm{n}} = n_{0} = 0.17$ $\mathrm{fm}^{-3}$.
We find that the mass splitting in the isospin doublet becomes larger as the asymmetry parameter increases.
Interestingly, the self-energies of the upper and lower components in the isospin doublet of $\bar{\mathrm{D}}$ meson are negative at $\delta = 1$, and they are still bound in neutron matter.
We may notice that the mass splitting increases linearly as a function of the asymmetry parameter $\delta$ when $|\delta|$ is small.
We can parametrize the self-energies of $\bar{\mathrm{D}}^{0}$ and $\mathrm{D}^{-}$ as 
\begin{eqnarray}
\Sigma_{\bar{\mathrm{D}}^{0}}^{(\mathrm{asym})}(n,\delta) &\simeq& \Sigma_{\bar{\mathrm{D}}}^{(\mathrm{sym})}(n) - \sigma_{\bar{\mathrm{D}}}(n) \delta, \\
\Sigma_{\mathrm{D}^{-}}^{(\mathrm{asym})}(n,\delta) &\simeq& \Sigma_{\bar{\mathrm{D}}}^{(\mathrm{sym})}(n) + \sigma_{\bar{\mathrm{D}}}(n) \delta,
\end{eqnarray}
for small $|\delta| \ll 1$, where $\Sigma_{\bar{\mathrm{D}}}^{(\mathrm{sym})}(n)$ is the self-energy of $\bar{\mathrm{D}}$ in isospin symmetric nuclear matter and $\pm \sigma_{\bar{\mathrm{D}}}(n) \delta$ denotes the deviation.
We obtain $\sigma_{\bar{\mathrm{D}}}(n)$ as a function of nuclear matter density $n$ in Fig.~\ref{fig:sigma_density_asy}.
We find $\sigma_{\bar{\mathrm{D}}}(n_0) = 30$ MeV at the normal nuclear matter density $n_{0}$.

In Fig.~\ref{fig:D_asymmetry}, we plot the results for $\mathrm{B}^{+}$ and $\mathrm{B}^{0}$ mesons also.
For the $\mathrm{B}$ meson case, the Fermi energy of neutrons or protons reaches at the critical Fermi momentum in Eq.~(\ref{eq:critical_Fermi}) at $|\delta| = 0.63$, and the self-energies of $\mathrm{B}^{+}$ and $\mathrm{B}^{0}$ mesons have imaginary parts for $|\delta| > 0.63$.
For small $|\delta|$, we define $\sigma_{\mathrm{B}}(n)$ for $\mathrm{B}$ meson by giving the self-energies of $\mathrm{B}^{+}$ and $\mathrm{B}^{0}$ mesons as
\begin{eqnarray}
\Sigma_{\mathrm{B}^{+}}^{(\mathrm{asym})}(n,\delta) &\simeq& \Sigma_{\mathrm{B}}^{(\mathrm{sym})}(n) - \sigma_{\mathrm{B}}(n) \delta, \\
\Sigma_{\mathrm{B}^{0}}^{(\mathrm{asym})}(n,\delta) &\simeq& \Sigma_{\mathrm{B}}^{(\mathrm{sym})}(n) + \sigma_{\mathrm{B}}(n) \delta,
\end{eqnarray}
where $\Sigma_{\mathrm{B}}^{(\mathrm{sym})}(n)$ is the self-energy of $\mathrm{B}$ meson in isospin symmetric nuclear matter.
We show $\sigma_{\mathrm{B}}(n)$ in Fig.~\ref{fig:sigma_density_asy}, and find $\sigma_{\mathrm{B}}(n_{0})=97$ MeV at the normal nuclear matter density $n_{0}$.

We notice again that the modifications of the self-energies of $\bar{\mathrm{D}}$ and $\mathrm{B}$ mesons are proportional to $\delta$.
This property can be contrasted with the change of the binding energy of a nucleon;
  the change of the binding energy per a nucleon is proportional to $\delta^{2}$ \cite{Kaiser:2001jx}.
Hence, for small deviation of isospin asymmetry ($|\delta| \ll 1$), the energy gain of the upper (lower) component of $\bar{\mathrm{D}}$ or $\mathrm{B}$ meson for $\delta >0$ ($\delta < 0$) overcomes the energy loss of the binding energy of a nucleon.

\begin{figure}[htbp]
\includegraphics[width=5cm,angle=-90]{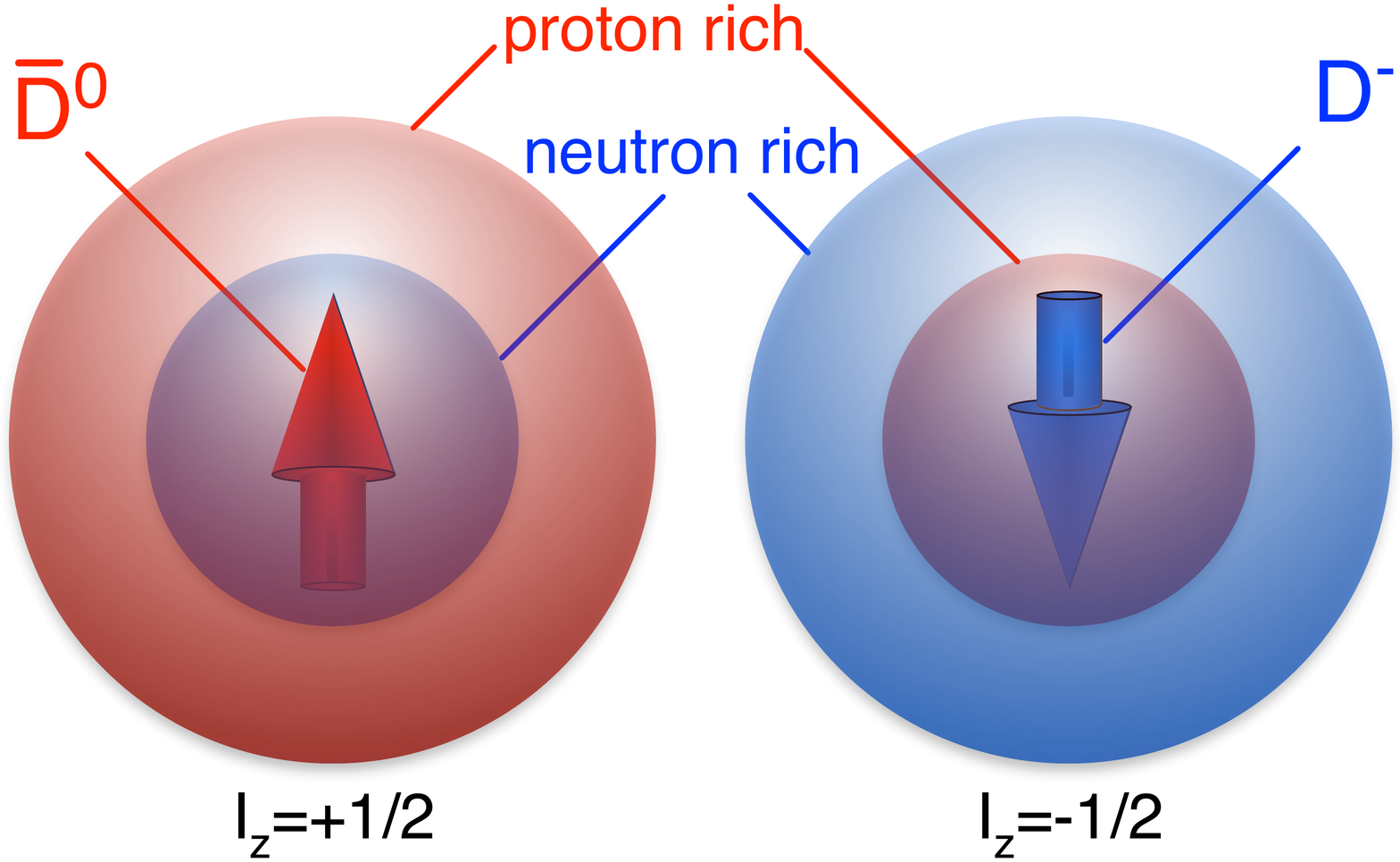}
\caption{Schematic picture of isospin density distribution around $\bar{\mathrm{D}}^{0}$ ($I_{\mathrm{z}}=+1/2$; left) and  $\mathrm{D}^{-}$ ($I_{\mathrm{z}}=-1/2$; right).}
\label{fig:Fig4_120612}
\end{figure}

The above result leads to an interesting phenomena about the spatial distribution of isospin density around $\bar{\mathrm{D}}$ or $\mathrm{B}$ meson in nuclear matter.
Let us suppose that $\bar{\mathrm{D}}$ or $\mathrm{B}$ meson with isospin $I_{\mathrm{z}}=+1/2$ (the upper component; $\bar{\mathrm{D}}^{0}$ or $\mathrm{B}^{+}$) is embedded in nuclear matter.
Here, we assume that the nuclear matter is uniformly isospin symmetric ($\delta = 0$) before $\bar{\mathrm{D}}$ or $\mathrm{B}$ meson is embedded.
From a view point of energy minimization,  
 it will occur that the isospin density becomes neutron-rich ($\delta > 0$) in the region near $\bar{\mathrm{D}}^{0}$ or $\mathrm{B}^{+}$ meson.
Because, 
 for small $|\delta|$,
 the energy gain by $\bar{\mathrm{D}}^{0}$ or $\mathrm{B}^{+}$ meson dominates against the energy loss of nuclear matter.
In order to conserve the total isospin in the whole nuclear matter, the region far apart from $\bar{\mathrm{D}}^{0}$ or $\mathrm{B}^{+}$ meson will become proton-rich, because that the region will be less affected by the existence of $\bar{\mathrm{D}}^{0}$ or $\mathrm{B}^{+}$ meson.
A similar phenomena would occur for $\bar{\mathrm{D}}$ or $\mathrm{B}$ meson with isospin $I_{\mathrm{z}}=-1/2$ (the lower component; $\mathrm{D}^{-}$ or $\mathrm{B}^{0}$), when the role of isospin up and down is just interchanged.
The spatial distribution of isospin density around $\bar{\mathrm{D}}$ or $\mathrm{B}$ meson embedded in uniform nuclear matter may be called ``isospin polarization."
Thus, the isospin of $\bar{\mathrm{D}}$ or $\mathrm{B}$ meson in nuclear matter would be dressed by the deformation of the isospin density by protons and neutrons.
We present the schematic picture in Fig.~\ref{fig:Fig4_120612}.
The self-energies in the previous section were obtained under the condition of the uniformity of the isospin density in the nuclear matter around  $\bar{\mathrm{D}}$ or $\mathrm{B}$ meson.
We expect that, if the variation of the isospin density in nuclear matter is taken into account, the binding energy of $\bar{\mathrm{D}}$ or $\mathrm{B}$ meson would become larger.
However, a quantitative discussion will be left for future studies.

The picture presented here may provide us with a new dynamics of $\bar{\mathrm{D}}$ or $\mathrm{B}$ meson in nuclear medium.
One can expect in general that 
the phenomena of ``isospin polarization" may occur if the hadron embedded in nuclear medium as impurity has the following two properties; (i) absence of strong decay
 and (ii) isospin non-singlet.
As for the first condition, 
we should note that $\mathrm{D}$ and $\bar{\mathrm{B}}$ mesons would have large decay widths
 which prevent to form stable states like isospin polarization in nuclear medium.
As for the second condition, we may consider in the strangeness sector that $\bar{\mathrm{K}}$ or $\mathrm{K}$ meson (isodoublet) can be embedded in nuclear medium \cite{Dote:2002db}.
However, it is known that systems of $\bar{\mathrm{K}}$ meson and nucleon are unstable because of the strong decays with large width and the nonnegligible absorption processes,
 although $\bar{\mathrm{K}}$ meson may be deeply bound in nuclear medium \footnote{See for example Ref.~\cite{Hyodo:2011ur} and the references therein.}.
  $\mathrm{K}$ meson is not bound in nuclear medium.
$\Sigma$, $\Sigma_{\mathrm{c}}$, and $\Sigma_{\mathrm{b}}$ baryons (isotriplet) will be also unstable in nuclear medium, because they have strong decays by $\pi$ emission and absorption processes with nucleons.
Therefore, we conclude that $\bar{\mathrm{D}}$ and $\mathrm{B}$ mesons are unique impurities which can exhibit
the stable isospin polarization in nuclear medium.

\section{Summary}

We discuss the mass modifications of $\bar{\mathrm{D}}$ and $\bar{\mathrm{D}}^{\ast}$ ($\mathrm{B}$ and $\mathrm{B}^{\ast}$) mesons in nuclear medium.
With respecting the heavy quark symmetry and chiral symmetry, we calculate the self-energies of $\bar{\mathrm{D}}$ and $\bar{\mathrm{D}}^{\ast}$ ($\mathrm{B}$ and $\mathrm{B}^{\ast}$) mesons by the pion exchange.
As results, we obtain the self-energies $-35.1$ MeV for $\bar{\mathrm{D}}$ meson and $-106.9$ MeV for $\mathrm{B}$ meson in isospin symmetric nuclear medium at the normal nuclear matter density $0.17$ $\mathrm{fm}^{-3}$.
Thus, both $\bar{\mathrm{D}}$ and $\mathrm{B}$ mesons can be bound in nuclear matter.
As an application to atomic nuclei, we find several energy levels of $\bar{\mathrm{D}}$ meson in $^{40}_{\bar{\mathrm{D}}}\mathrm{Ca}$ and $^{208}_{\bar{\mathrm{D}}}\mathrm{Pb}$ nuclei, in which a $\bar{\mathrm{D}}$ meson is bound in $^{40}\mathrm{Ca}$ and $^{208}\mathrm{Pb}$ nuclei.
We obtain the self-energies $-153.9-i170.1$ MeV for $\bar{\mathrm{D}}^{\ast}$ meson and $-203.3-i117.5$ MeV for $\mathrm{B}^{\ast}$ meson at the normal nuclear matter density.
However, due to their large imaginary parts, the individual energy levels of $\bar{\mathrm{D}}^{\ast}$ and $\mathrm{B}^{\ast}$ in atomic nuclei will not be seen clearly.
We further discuss the isospin asymmetric nuclear matter, and find that the upper and lower components in the isospin doublet $\bar{\mathrm{D}}$ or $\mathrm{B}$ meson split in mass for any small isospin asymmetry.
With this result, we present the possible phenomena of ``isospin polarization" induced by $\bar{\mathrm{D}}$ or $\mathrm{B}$ meson in nuclear matter.

For future studies, we will need to investigate the following several effects, which are omitted in the present formalism.
The discussions about the higher order effects of the pion exchange, the sigma term, $\Delta$ excitations, and so on, will be required for an extension to higher densities \cite{Fritsch:2004nx,Fiorilla:2011qr}.
It will be also necessary to consider the modification of the QCD vacuum at finite density.
In experiments the nuclei with $\bar{\mathrm{D}}$ or $\mathrm{B}$ meson may be produced in collisions of hadron (antiproton, $\pi$) to nucleus \cite{Sibirtsev:1999js,Golubeva:2002au}, in nucleus-nucleus collisions \cite{Cassing:2000vx} and in relativistic heavy ion collisions \cite{Cho:2010db,Cho:2011ew}.
The future experimental studies in accelerator facilities, such as J-PARC, FAIR at GSI, RHIC, LHC, and so on, will be expected to be performed.

\section*{Acknowledgments}
This work is supported in part by Grant-in-Aid for Scientific Research on 
Priority Areas ``Elucidation of New Hadrons with a Variety of Flavors 
(E01: 21105006)" (S.Y.) and by ``Grant-in-Aid for Young Scientists (B)
22740174" (K.S.), from 
the ministry of Education, Culture, Sports, Science and Technology of
Japan.

\end{document}